\documentclass[twocolumn,letterpaper,aps,prd,longbibliography,superscriptaddress,nofootinbib,floatfix]{revtex4-2}

\usepackage{graphicx}	% Include figure files
\usepackage{amsmath}
\usepackage[utf8]{inputenc} % include the sivers bibtex
\usepackage[pdftex]{hyperref} 
\usepackage{dcolumn}	% Align table columns on decimal point
\usepackage{xspace}	% Include xspace

\newcommand{\nch}{\mbox{$N_{\rm ch}$}\xspace}
\newcommand{\NCH}{N_{\rm ch}}
\newcommand{\nchach}{\mbox{$N_{\rm ch}/\langle{N_{\rm ch}}\rangle$}\xspace}

\newcommand{\sqrts}{\mbox{$\sqrt{s}$}\xspace}

\newcommand{\jpsi}{\mbox{$J/\psi$}\xspace}

\newcommand{\psipp}{\mbox{$\psi(2S)$}\xspace}

\newcommand{\pp}{\mbox{$p$+$p$}\xspace}

\newcommand{\pythia}{\mbox{\sc pythia8}\xspace}

\begin{document}

\title{Multiplicity dependent $J/\psi$ and $\psi(2S)$ production at 
forward and backward rapidity in $p$$+$$p$ collisions at $\sqrt{s}=200$ GeV}

%\author{PHENIX Collaboration Author List (Brant will insert later) }

\newcommand{\abilene}{Abilene Christian University, Abilene, Texas 79699, USA}
\newcommand{\augie}{Department of Physics, Augustana University, Sioux Falls, South Dakota 57197, USA}
\newcommand{\banaras}{Department of Physics, Banaras Hindu University, Varanasi 221005, India}
\newcommand{\barc}{Bhabha Atomic Research Centre, Bombay 400 085, India}
\newcommand{\baruch}{Baruch College, City University of New York, New York, New York, 10010 USA}
\newcommand{\bnlcoll}{Collider-Accelerator Department, Brookhaven National Laboratory, Upton, New York 11973-5000, USA}
\newcommand{\bnlphys}{Physics Department, Brookhaven National Laboratory, Upton, New York 11973-5000, USA}
\newcommand{\caucr}{University of California-Riverside, Riverside, California 92521, USA}
\newcommand{\charlesczech}{Charles University, Faculty of Mathematics and Physics, 180 00 Troja, Prague, Czech Republic}
\newcommand{\cns}{Center for Nuclear Study, Graduate School of Science, University of Tokyo, 7-3-1 Hongo, Bunkyo, Tokyo 113-0033, Japan}
\newcommand{\colorado}{University of Colorado, Boulder, Colorado 80309, USA}
\newcommand{\columbia}{Columbia University, New York, New York 10027 and Nevis Laboratories, Irvington, New York 10533, USA}
\newcommand{\czechtech}{Czech Technical University, Zikova 4, 166 36 Prague 6, Czech Republic}
\newcommand{\debrecen}{Debrecen University, H-4010 Debrecen, Egyetem t{\'e}r 1, Hungary}
\newcommand{\elte}{ELTE, E{\"o}tv{\"o}s Lor{\'a}nd University, H-1117 Budapest, P{\'a}zm{\'a}ny P.~s.~1/A, Hungary}
\newcommand{\ewha}{Ewha Womans University, Seoul 120-750, Korea}
\newcommand{\fsu}{Florida State University, Tallahassee, Florida 32306, USA}
\newcommand{\gsu}{Georgia State University, Atlanta, Georgia 30303, USA}
\newcommand{\hiroshima}{Physics Program and International Institute for Sustainability with Knotted Chiral Meta Matter (SKCM2), Hiroshima University, Higashi-Hiroshima, Hiroshima 739-8526, Japan}
\newcommand{\howard}{Department of Physics and Astronomy, Howard University, Washington, DC 20059, USA}
\newcommand{\hunrenatomki}{HUN-REN ATOMKI, H-4026 Debrecen, Bem t{\'e}r 18/c, Hungary}
\newcommand{\ihepprot}{IHEP Protvino, State Research Center of Russian Federation, Institute for High Energy Physics, Protvino, 142281, Russia}
\newcommand{\illuiuc}{University of Illinois at Urbana-Champaign, Urbana, Illinois 61801, USA}
\newcommand{\inrras}{Institute for Nuclear Research of the Russian Academy of Sciences, prospekt 60-letiya Oktyabrya 7a, Moscow 117312, Russia}
\newcommand{\instpasczech}{Institute of Physics, Academy of Sciences of the Czech Republic, Na Slovance 2, 182 21 Prague 8, Czech Republic}
\newcommand{\isu}{Iowa State University, Ames, Iowa 50011, USA}
\newcommand{\jaea}{Advanced Science Research Center, Japan Atomic Energy Agency, 2-4 Shirakata Shirane, Tokai-mura, Naka-gun, Ibaraki-ken 319-1195, Japan}
\newcommand{\jeonbuk}{Jeonbuk National University, Jeonju, 54896, Korea}
\newcommand{\kek}{KEK, High Energy Accelerator Research Organization, Tsukuba, Ibaraki 305-0801, Japan}
\newcommand{\korea}{Korea University, Seoul 02841, Korea}
\newcommand{\kurchatov}{National Research Center ``Kurchatov Institute", Moscow, 123098 Russia}
\newcommand{\kyoto}{Kyoto University, Kyoto 606-8502, Japan}
\newcommand{\lawllnl}{Lawrence Livermore National Laboratory, Livermore, California 94550, USA}
\newcommand{\losalamos}{Los Alamos National Laboratory, Los Alamos, New Mexico 87545, USA}
\newcommand{\lund}{Department of Physics, Lund University, Box 118, SE-221 00 Lund, Sweden}
\newcommand{\lyon}{IPNL, CNRS/IN2P3, Univ Lyon, Universit{\'e} Lyon 1, F-69622, Villeurbanne, France}
\newcommand{\maryland}{University of Maryland, College Park, Maryland 20742, USA}
\newcommand{\mass}{Department of Physics, University of Massachusetts, Amherst, Massachusetts 01003-9337, USA}
\newcommand{\mate}{MATE, Institute of Technology, Laboratory of Femtoscopy, K\'aroly R\'obert Campus, H-3200 Gy\"ongy\"os, M\'atrai \'ut 36, Hungary}
\newcommand{\michigan}{Department of Physics, University of Michigan, Ann Arbor, Michigan 48109-1040, USA}
\newcommand{\miss}{Mississippi State University, Mississippi State, Mississippi 39762, USA}
\newcommand{\muhlenberg}{Muhlenberg College, Allentown, Pennsylvania 18104-5586, USA}
\newcommand{\nara}{Nara Women's University, Kita-uoya Nishi-machi Nara 630-8506, Japan}
\newcommand{\natmephi}{National Research Nuclear University, MEPhI, Moscow Engineering Physics Institute, Moscow, 115409, Russia}
\newcommand{\newmex}{University of New Mexico, Albuquerque, New Mexico 87131, USA}
\newcommand{\nmsu}{New Mexico State University, Las Cruces, New Mexico 88003, USA}
\newcommand{\northcg}{Physics and Astronomy Department, University of North Carolina at Greensboro, Greensboro, North Carolina 27412, USA}
\newcommand{\ohio}{Department of Physics and Astronomy, Ohio University, Athens, Ohio 45701, USA}
\newcommand{\ornl}{Oak Ridge National Laboratory, Oak Ridge, Tennessee 37831, USA}
\newcommand{\orsay}{IPN-Orsay, Univ.~Paris-Sud, CNRS/IN2P3, Universit\'e Paris-Saclay, BP1, F-91406, Orsay, France}
\newcommand{\peking}{Peking University, Beijing 100871, People's Republic of China}
\newcommand{\pnpi}{PNPI, Petersburg Nuclear Physics Institute, Gatchina, Leningrad region, 188300, Russia}
\newcommand{\riken}{RIKEN Nishina Center for Accelerator-Based Science, Wako, Saitama 351-0198, Japan}
\newcommand{\rikjrbrc}{RIKEN BNL Research Center, Brookhaven National Laboratory, Upton, New York 11973-5000, USA}
\newcommand{\rikkyo}{Physics Department, Rikkyo University, 3-34-1 Nishi-Ikebukuro, Toshima, Tokyo 171-8501, Japan}
\newcommand{\saispbstu}{Saint Petersburg State Polytechnic University, St.~Petersburg, 195251 Russia}
\newcommand{\seoulnat}{Department of Physics and Astronomy, Seoul National University, Seoul 151-742, Korea}
\newcommand{\stonybrkc}{Chemistry Department, Stony Brook University, SUNY, Stony Brook, New York 11794-3400, USA}
\newcommand{\stonycrkp}{Department of Physics and Astronomy, Stony Brook University, SUNY, Stony Brook, New York 11794-3800, USA}
\newcommand{\tenn}{University of Tennessee, Knoxville, Tennessee 37996, USA}
\newcommand{\titech}{Department of Physics, Tokyo Institute of Technology, Oh-okayama, Meguro, Tokyo 152-8551, Japan}
\newcommand{\tsukuba}{Tomonaga Center for the History of the Universe, University of Tsukuba, Tsukuba, Ibaraki 305, Japan}
\newcommand{\usmma}{United States Merchant Marine Academy, Kings Point, New York 11024, USA}
\newcommand{\vandy}{Vanderbilt University, Nashville, Tennessee 37235, USA}
\newcommand{\weizmann}{Weizmann Institute, Rehovot 76100, Israel}
\newcommand{\wigner}{Institute for Particle and Nuclear Physics, HUN-REN Wigner Research Centre for Physics, (HUN-REN Wigner RCP, RMI), H-1525 Budapest 114, POBox 49, Budapest, Hungary}
\newcommand{\yonsei}{Yonsei University, IPAP, Seoul 120-749, Korea}
\newcommand{\zagreb}{Department of Physics, Faculty of Science, University of Zagreb, Bijeni\v{c}ka c.~32 HR-10002 Zagreb, Croatia}
\newcommand{\zambia}{Department of Physics, School of Natural Sciences, University of Zambia, Great East Road Campus, Box 32379, Lusaka, Zambia}
\affiliation{\abilene}
\affiliation{\augie}
\affiliation{\banaras}
\affiliation{\barc}
\affiliation{\baruch}
\affiliation{\bnlcoll}
\affiliation{\bnlphys}
\affiliation{\caucr}
\affiliation{\charlesczech}
\affiliation{\cns}
\affiliation{\colorado}
\affiliation{\columbia}
\affiliation{\czechtech}
\affiliation{\debrecen}
\affiliation{\elte}
\affiliation{\ewha}
\affiliation{\fsu}
\affiliation{\gsu}
\affiliation{\hiroshima}
\affiliation{\howard}
\affiliation{\hunrenatomki}
\affiliation{\ihepprot}
\affiliation{\illuiuc}
\affiliation{\inrras}
\affiliation{\instpasczech}
\affiliation{\isu}
\affiliation{\jaea}
\affiliation{\jeonbuk}
\affiliation{\kek}
\affiliation{\korea}
\affiliation{\kurchatov}
\affiliation{\kyoto}
\affiliation{\lawllnl}
\affiliation{\losalamos}
\affiliation{\lund}
\affiliation{\lyon}
\affiliation{\maryland}
\affiliation{\mass}
\affiliation{\mate}
\affiliation{\michigan}
\affiliation{\miss}
\affiliation{\muhlenberg}
\affiliation{\nara}
\affiliation{\natmephi}
\affiliation{\newmex}
\affiliation{\nmsu}
\affiliation{\northcg}
\affiliation{\ohio}
\affiliation{\ornl}
\affiliation{\orsay}
\affiliation{\peking}
\affiliation{\pnpi}
\affiliation{\riken}
\affiliation{\rikjrbrc}
\affiliation{\rikkyo}
\affiliation{\saispbstu}
\affiliation{\seoulnat}
\affiliation{\stonybrkc}
\affiliation{\stonycrkp}
\affiliation{\tenn}
\affiliation{\titech}
\affiliation{\tsukuba}
\affiliation{\usmma}
\affiliation{\vandy}
\affiliation{\weizmann}
\affiliation{\wigner}
\affiliation{\yonsei}
\affiliation{\zagreb}
\affiliation{\zambia}
\author{N.J.~Abdulameer} \affiliation{\debrecen} \affiliation{\hunrenatomki}
\author{U.~Acharya} \affiliation{\gsu}
\author{C.~Aidala} \affiliation{\michigan} 
\author{Y.~Akiba} \email[PHENIX Spokesperson: ]{akiba@rcf.rhic.bnl.gov} \affiliation{\riken} \affiliation{\rikjrbrc}
\author{M.~Alfred} \affiliation{\howard} 
\author{V.~Andrieux} \affiliation{\michigan} 
\author{S.~Antsupov} \affiliation{\saispbstu}
\author{N.~Apadula} \affiliation{\isu} 
\author{H.~Asano} \affiliation{\kyoto} \affiliation{\riken} 
\author{B.~Azmoun} \affiliation{\bnlphys} 
% ...
\collaboration{PHENIX Collaboration}  \noaffiliation

\date{\today}

%-----------------------------------------------------------------------------|

\begin{abstract}

%\linenumbers

The $J/\psi$ and $\psi(2S)$ charmonium states, composed of $c\bar{c}$ 
quark pairs and known since the 1970s, are widely believed to serve as 
ideal probes to test quantum chromodynamics in high-energy hadronic 
interactions.  However, there is not yet a complete understanding 
of the charmonium-production mechanism.  Recent measurements of 
$J/\psi$ production as a function of event charged-particle 
multiplicity at the collision energies of both the Large Hadron 
Collider (LHC) and the Relativistic Heavy Ion Collider (RHIC) show 
enhanced $J/\psi$ production yields with increasing multiplicity. One 
potential explanation for this type of dependence is multiparton 
interactions (MPI).  We carry out the first measurements of 
self-normalized $J/\psi$ yields and the $\psi(2S)$ to $J/\psi$ ratio at 
both forward and backward rapidities as a function of self-normalized 
charged-particle multiplicity in $p$$+$$p$ collisions at $\sqrt{s}=200$ 
GeV. In addition, detailed {\sc pythia} studies tuned to RHIC energies 
were performed to investigate the MPI impacts. We find that the PHENIX 
data at RHIC are consistent with recent LHC measurements and can only 
be described by {\sc pythia} calculations that include MPI effects. The 
forward and backward $\psi(2S)$ to $J/\psi$ ratio, which serves as a 
unique and powerful approach to study final-state effects on charmonium 
production, is found to be less dependent on the charged-particle 
multiplicity.

\end{abstract}

\maketitle

%\textbf{*** page break for PRD Letters word count ***}  
%\clearpage

%========================================
% Introduction (Shi:2023gnw, Aguilar:2021sfa, Oh:2023lvj)
%========================================
% motivation

Charmonium, a bound $c\bar{c}$ state, has been studied extensively over 
the past several decades, but a clear understanding of its formation 
has not yet been reached.  Several models are currently available to 
describe the evolution of a $c\bar{c}$ pair into the bound \jpsi or 
\psipp meson, such as the nonrelativistic-quantum-chromodynamics 
(NRQCD)~\cite{Brambilla:1999xf}, color-evaporation~\cite{Amundson:1996qr}, 
color-singlet~\cite{Kuhn:1979zb}, and 
jet-fragmentation~\cite{Baumgart:2014upa} models.  
The formation appears to involve both 
perturbative (above the $\Lambda_{\rm QCD}$ scale) and nonperturbative 
(below the $\Lambda_{\rm QCD}$ scale) aspects of QCD.  The initial 
creation of $c\bar c$ pairs through hard scattering can be described as 
perturbative, and their evolution into a color-neutral state is likely 
nonperturbative.

In this Letter, we present PHENIX measurements at the Relativistic 
Heavy Ion Collider (RHIC) of the self-normalized \jpsi production 
versus self-normalized event multiplicity to study the 
multi-parton-interaction (MPI) effects at forward rapidity as well as 
the \psipp to \jpsi ratio, which is sensitive to final-state 
interactions in \pp collisions at RHIC energies.  We measure inclusive 
\jpsi and \psipp production without separating prompt from nonprompt 
charmonium because the nonprompt contributions are less than 3\% of the 
total charmonium production.

The STAR Experiment at RHIC has measured the self-normalized \jpsi 
yields as a function of self-normalized charged-particle multiplicity 
at midrapidity in \pp collisions~\cite{STAR:2018smh}.  The results show 
an increase in the \jpsi yields with increasing multiplicity, a 
dependence suggesting MPI. Additionally, at Large-Hadron-Collider (LHC) 
energies, the ALICE experiment has reported similar results for the 
normalized \jpsi yields versus event multiplicity at both 
forward rapidity~\cite{ALICE:2011gej,ALICE:2021zkd} and 
midrapidity~\cite{ALICE:2020msa}. In addition to potential MPI, the 
\psipp to \jpsi ratio could reveal final-state effects from either hot 
(formation of quark-gluon 
plasma~\cite{BRAHMS:2004adc,PHENIX:2004vcz,PHOBOS:2004zne,STAR:2005gfr}), 
or cold (comover-interaction model~\cite{Ferreiro:2014bia}) 
nuclear-matter effects. In early 2024, the LHCb collaboration reported 
the ratio of the normalized \psipp to \jpsi in \pp collisions at 
$\sqrts=13$~TeV as a function of multiplicity~\cite{LHCb:2023xie}, 
where suppression of promptly produced charmonia is observed at high 
event multiplicity.

%========================================
% Data Sets & Experimental Setup (PHENIX-AnaNote-1490, 1498)
%========================================
% data set
%---------------------------------------

%========================================
% experimental setup
%---------------------------------------

The present analysis relies on the data from the PHENIX 
experiment~\cite{PHENIX:1998vmi} obtained using the muon-arms detector 
subsystem covering $1.2<|\eta|<2.4$, which includes the muon tracker 
(MuTr), the muon identifier (MuID), the hadron absorbers, the forward 
silicon vertex detectors 
(FVTX)~\cite{Akikawa:2003zs,Adachi:2013qha,Aidala:2013vna,Allen:2003zt} 
in the forward rapidity region and the central arm barrel silicon 
vertex tracker (VTX)~\cite{Nouicer:2007rb} at $|\eta|<1.0$. Two 
beam-beam counters (BBC), covering the full azimuth and 
$3.1<|\eta|<3.9$, measure the vertex position along the beamline, 
located at $z={\pm}144$~cm from the nominal interaction point. The 
BBCs also serve as the minimum-bias (MB) trigger and measure the beam 
luminosity.

The \pp data set used in this analysis was collected in 2015 and 
recorded at $\sqrts=200$~GeV center-of-mass energy. The analyzed events 
were selected by the MB trigger and the dimuon triggers, which  
required two or more muon tracks in the MuID. 
The collision vertex was constrained to be within ${\pm}10$~cm with 
respect to the center of the interaction region.  The total sampled 
luminosity for the \pp data set is 47 pb$^{-1}$.

%========================================
% Data Analysis (PHENIX-AnaNote-1490, 1498)
%========================================
% observable definitions
%---------------------------------------

The observable used in this analysis has theoretical~\cite{Ferreiro:2012} 
and experimental~\cite{ALICE:2012165} motivations. We first define the 
self-normalized event charged-particle multiplicity \nchach, where \nch 
represents the number of reconstructed charged-particle tracks detected by 
the forward- or backward-rapidity FVTX with $1.2<\eta<2.4$ and 
$-2.4<\eta<-1.2$, and $\langle\NCH\rangle$ represents the average \nch for 
MB events. Then the relative yield of \jpsi, denoted as 
$N_{J/\psi}/{\langle}N_{J/\psi}\rangle$ in a given \nchach range, is 
measured by the forward or backward muon arms covering $1.2<y<2.2$ and 
$-2.2<y<-1.2$, respectively:

\begin{equation}
\label{eq:rel_num}
    N_{J/\psi}/\langle N_{J/\psi} \rangle = \frac{N_{J/\psi}^{\rm raw}}
{N_{\rm evt} } \frac{ N_{\rm evt}^{\rm total} }{N_{J/\psi}^{\rm raw,total}} 
\frac{\varepsilon^{\rm MB}_{\rm trig}}
{\varepsilon^{J/\psi}_{\rm trig}} 
\frac{\langle \varepsilon^{J/\psi}_{\rm trig}\rangle}
{\langle\varepsilon^{\rm MB}_{\rm trig}\rangle} f_{\rm coll},
\end{equation}

\noindent where $N_{J/\psi}^{\rm raw}$ is the raw number of \jpsi 
signal yield extracted from the dimuon invariant mass fit, and $N_{\rm 
evt}$ is the number of recorded MB events for a certain \nch bin. The 
average MB efficiency is $\langle\varepsilon^{\rm MB}_{\rm 
trig}\rangle=55{\pm}5$\% and the $J/\psi$ efficiency is 
$\langle\varepsilon^{J/\psi}_{\rm trig}\rangle=79{\pm}2$\%. The 
superscript ``total" stands for quantities integrated over all \nch. 
The $\epsilon^{\rm MB}_{\rm trig}$ is the MB trigger efficiency, 
$\epsilon^{J/\psi}_{\rm trig}$ is the dimuon trigger efficiency. The 
$f_{\rm coll}$ is a correction factor for multiple collisions. The same 
observable is then measured for \psipp, and the ratio of relative 
yields is defined as 
$({N_{\psi(2S)}/ N_{J/\psi})/\langle{N_{\psi(2S)}}/N_{J/\psi}\rangle}$.
We present the data in terms of the same arms defined as measuring 
\nch and $N_{J/\psi}$ at the same rapidity range, and opposite arms 
defined as measuring \nch and $N_{J/\psi}$ at the opposite rapidity 
range. When combining the data from two arms, we define 
$1.2<|\eta|<2.4$ for \nch and $1.2<|y|<2.2$ for $N_{J/\psi}$.

%========================================
% trigger efficiency
%---------------------------------------

In PHENIX the trigger efficiency has event-multiplicity dependence due 
to finite acceptance of the BBC. The event multiplicity-dependent MB 
and dimuon trigger efficiencies are determined using a data-driven 
approach. The MB trigger efficiency versus event multiplicity was 
evaluated with random-collision-clock triggered events by checking 
whether the MB trigger is fired for events satisfying the MB trigger 
condition. To check the multiplicity dependence of the MB trigger for 
hard-scattering events, we require at least one EMCAL cluster of $E>2$ 
GeV on top of the random-collision-clock trigger

%========================================
% multiple collisions estimate
%---------------------------------------

The possibility of multiple-collision events, defined as an event 
having \pp collisions in addition to the one that produced the \jpsi 
or \psipp, increases with increasing multiplicity. A data-driven method 
is utilized to estimate the fraction of multiple collision events, 
which assumes Poisson statistics and estimates the BBC (MB) trigger 
rate $R_{\rm BBC}$ using the following formula:

\begin{equation}
R_{\rm BBC} = f_{BC}[1-e^{-\mu\epsilon_F}-e^{-\mu\epsilon_B}-e^{-\mu(\epsilon_F+\epsilon_B+\epsilon_{FB})}],
\end{equation}

\noindent where $f_{BC}$ is the beam-crossing rate at RHIC, $\mu$ is the mean 
number of collisions, $\epsilon_F$ ($\epsilon_B$) is the trigger efficiency 
($\approx$75\%) of each BBC at forward (backward) rapidity, and 
$\epsilon_{FB}$ is the trigger efficiency ($\approx$50\%) when both BBCs are 
fired. The maximum $R_{\rm BBC}$ during the \pp data run is 2.5 MHz, 
corresponding to an $\approx$10\% rate of double collision events. Due to the 
average $R_{\rm BBC}$ being $\approx$1 MHz and the low probability of having 
more than two collisions per beam crossing, only contributions from double 
collisions were considered. The ratio of double to single collisions, 
$f_{\rm coll}$, is evaluated as a function of the measured-BBC-trigger rate 
$R_{\rm BBC}$.

%========================================
% mass fits, signal
%---------------------------------------

%-------------------------------------------------------- Fig_1
\begin{figure*}[htb]
\begin{minipage}{0.48\linewidth}
\includegraphics[width=0.99\linewidth]{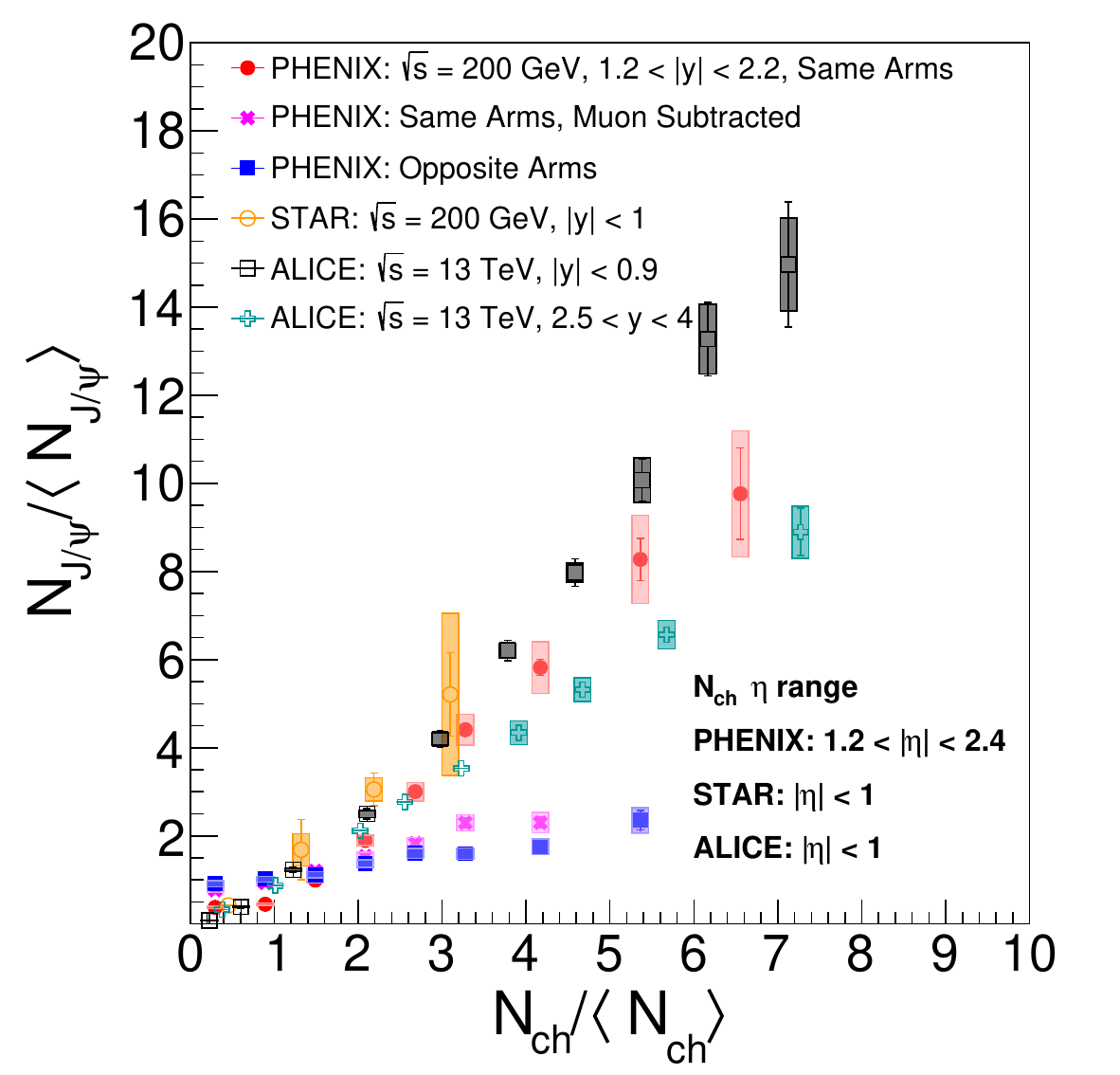}
\caption{\label{fig:jpsi_before_after}
The self-normalized $J/\psi$ yield as a function of self-normalized 
charged-particle multiplicity in \pp collisions at various energies. 
The PHENIX data combine forward and backward $J/\psi$ with forward 
and backward FVTX, respectively. The contribution of $J/\psi$ daughter-muon 
tracks included in the charged-particle multiplicity (solid circle) and 
the ones subtracting the $J/\psi$ daughter-muon-tracks contribution 
(solid cross) in the same arms as well as the ones with opposite arm 
(solid square) are presented. The PHENIX measurements are compared with 
ALICE measurements at forward rapidity~\cite{ALICE:2021zkd} and 
at midrapidity~\cite{ALICE:2020msa}, and STAR~\cite{STAR:2018smh} 
measurements at midrapidity.  The error bars (boxes) denote the 
statistical (systematic) uncertainties.
}
\end{minipage}
%\end{figure}
%-------------------------------------------------------- Fig_2
%\begin{figure}[tbh]
\hspace{0.2cm}
\vspace{0.8cm}
\begin{minipage}{0.48\linewidth}
\vspace{-1.1cm}
\includegraphics[width=0.99\linewidth]{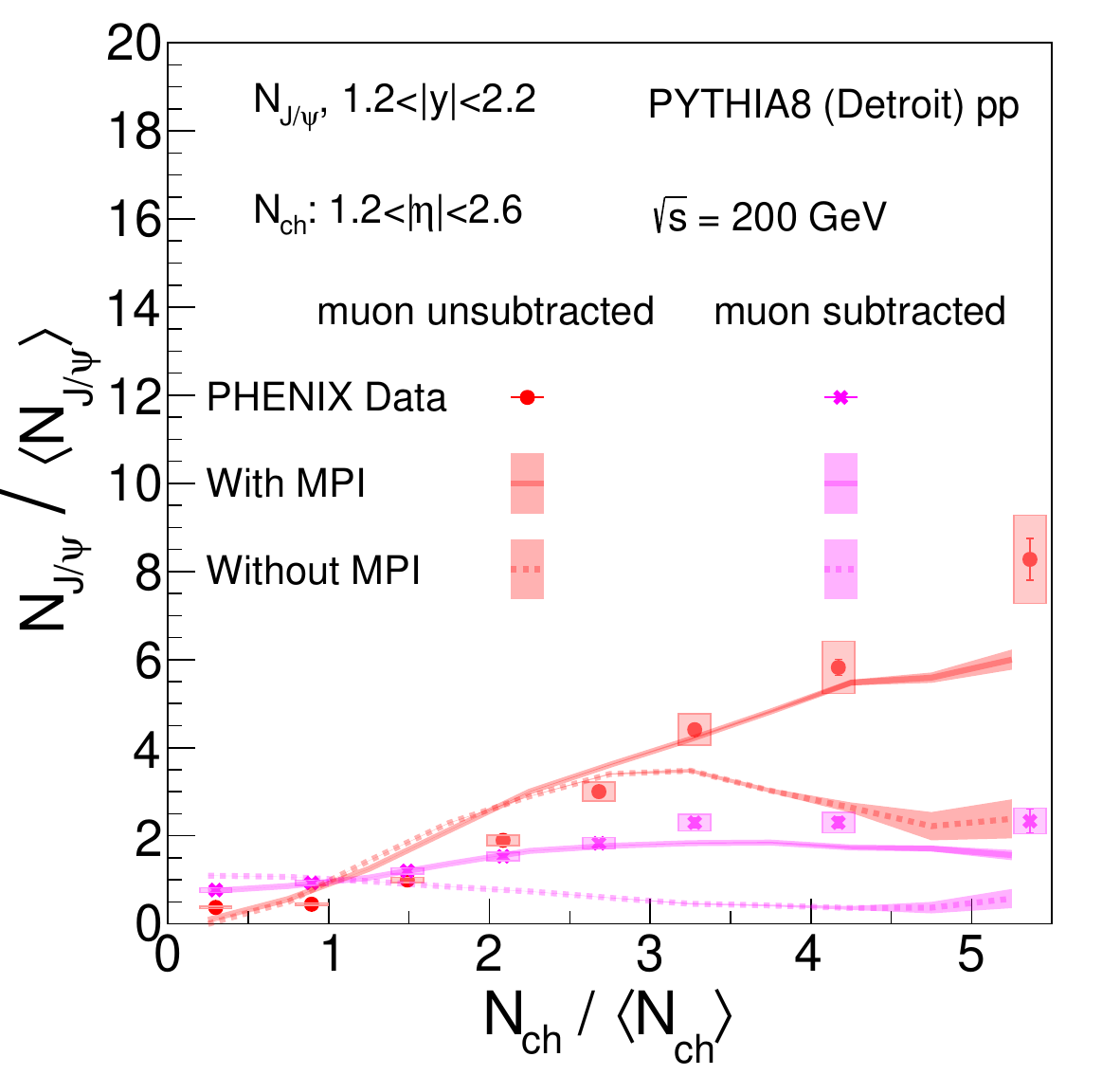}
\caption{\label{fig:jpsi_pythia}
The self-normalized \jpsi yields are shown as a function of
self-normalized charged-particle multiplicity in \pp collisions at
$\sqrt{s}=200$~GeV. The solid circle and solid cross data points,
respectively, represent the measurements before and after the \jpsi muon
tracks have been subtracted (see text for more details). The error bars
(boxes) denote the statistical (systematic) uncertainties. Data are
compared to \pythia Detroit tuned calculations in the pseudorapidity
range of $2.1 < |\eta| < 2.6$ for charged particles with (solid lines)
and without (dash lines) considering MPI effects~\cite{Oh:2023lvj}.
}
\end{minipage}
\end{figure*}

A crystal-ball (CB) function~\cite{Gaiser:1982yw} is used to model the 
\jpsi and \psipp signal shapes to extract the raw yields from the 
dimuon invariant-mass distribution. The tail parameters $\alpha$ and 
$n$ of the CB function were fixed with values from integrated 
multiplicity data for \jpsi and simulation studies for $\psi(2S)$. The 
FVTX detector delivers the mass resolution needed for \psipp 
measurements but reduces the acceptance due to a finite matching 
efficiency between the MuTr and FVTX detectors. For 
$N_{J/\psi}/\langle{N_{J/\psi}}\rangle$
 as a function of $\NCH/\langle\NCH\rangle$ 
shown in Figs.~\ref{fig:jpsi_before_after} and~\ref{fig:jpsi_pythia}, 
the FVTX detector is not used to maximize \jpsi signal statistics. For 
\psipp measurements with the FVTX detector, a Gaussian function is used 
alongside the CB function to account for the broadening of signal shape 
due to misassociation between the FVTX and MuTr 
detectors~\cite{PHENIX:2016vmz}. The individual multiplicity bins are 
fitted by fixing the shape parameters of the CB function to the 
integrated multiplicity fit results to extract the $J/\psi$ signal 
yield.

For the \jpsi analysis without the FVTX detector, an exponential decay 
function is used to describe the background. For the $\psi(2S)$ to $J/\psi$ 
ratio, a modified Hagedorn function is used to model the combinatorial 
and correlated-background 
contributions~\cite{Aidala:2018ajl,PHENIX:2022nrm}. The shape of the 
combinatorial background is estimated with real data using the 
mixed-event method, where opposite sign single muons are selected from 
different events. The combinatorial background normalization is 
obtained from like-sign single muons selected from the same events. The 
correlated background shape is determined using detailed simulation 
studies~\cite{Aidala:2018ajl} and is implemented in the total fit 
function by fixing three of the five parameters of the modified 
Hagedorn function to constrain the shape.

%========================================
% Systematic Uncertainty
%========================================
% AN 1490

The sources of systematic uncertainties for the \jpsi measurements 
include the following: the \jpsi reconstruction efficiency, the \jpsi 
trigger efficiency, the MB trigger efficiency, and the 
multiple-collision-correction factor $f_{\rm coll}$ per beam crossing. 
A systematic uncertainty is assigned for the \jpsi reconstruction 
efficiency based on the largest variation when three different 
$z$-vertex selections are applied. For the systematic uncertainty 
related to the \jpsi trigger efficiency, the efficiency as a function 
of \nch is first fit with the following function: $f(x) = p_0 + 
p_1e^{-p_2x}$. Then, the efficiency distribution is re-evaluated by 
moving the parameter values by $\pm1\sigma$ of the statistical 
uncertainty in the fit result. The largest difference with respect to 
the nominal value is assigned as the systematic uncertainty. The 
MB-trigger efficiency can be affected by multiple collisions. The 
systematic uncertainty related to the MB-trigger efficiency is 
determined by dividing the trigger efficiency into three different \pp 
collision rates, determined by the MB trigger: 600--800, 1000--1500, 
and 2000--2500 kHz. The trigger efficiency for the high and low trigger 
rates are compared to the central rate, and the maximum deviation is 
scaled by $1/\sqrt{12}$ (the standard deviation of a uniform 
distribution), which is assigned as the systematic uncertainty.  The 
systematic uncertainty related to the multiple-collision correction 
factor is evaluated by comparing correction factors using an 
alternate method, which is the ratio of probability distributions 
$P(\NCH)$ between two different trigger rates, less than 500 and 
1000--1500 kHz.

% AN 1498

The sources of systematic uncertainty for the normalized \psipp to 
\jpsi ratio include the following: the mixed-event background 
normalization, the signal shapes (fixed CB tail parameters and the 
second Gaussian width), and the correlated background shape.  Most 
systematic uncertainties cancel for the double ratio, including 
uncertainties related to fixing the CB tail parameters, the second 
Gaussian width, and the multiple collision correction factor. The 
correlated background uncertainty was determined by allowing two of the 
five parameters in the fit function to vary and comparing the resulting 
signal yields from each fit.  The determination of the uncertainty in 
the normalization of the mixed-event background follows the methods 
described in Ref.~\cite{PHENIX:2016vmz} and an uncertainty is assigned by 
repeating the fit to the invariant-mass distribution over a mass range 
extended first below and then above the nominal mass range of the fit.

%========================================
% Sanghoon Simulation Study (Oh:2023lvj)
%========================================

%========================================
%  Results
%========================================

%-------------------------------------------------------- Fig_3
\begin{figure*}[htb]  
\includegraphics[width=0.99\linewidth]{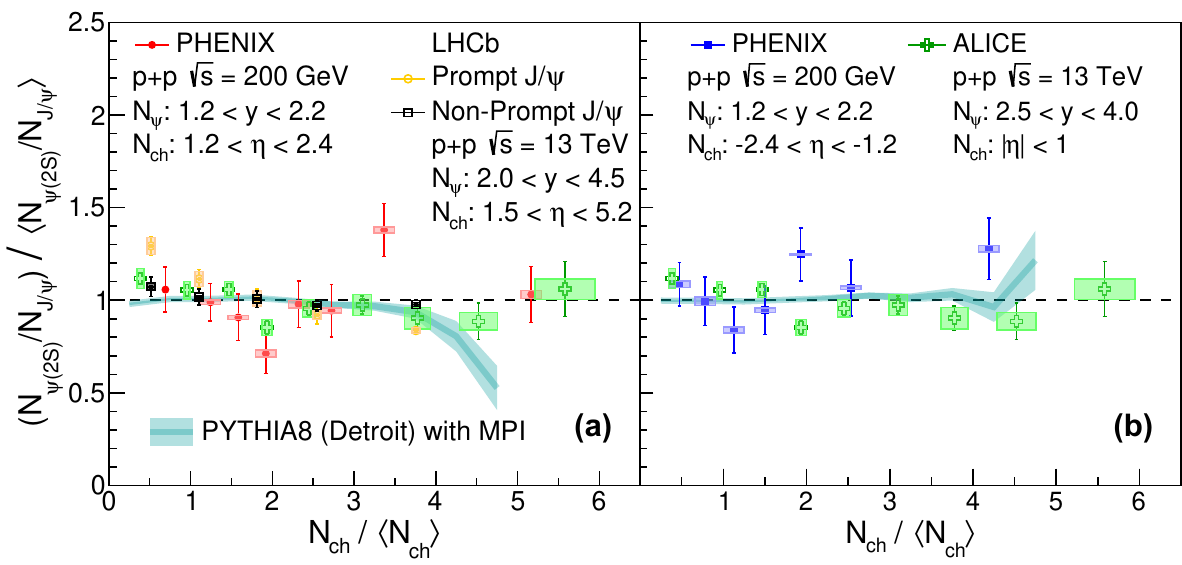}
\caption{\label{fig:psip_alice_pythia} The self-normalized \psipp to 
\jpsi ratio at forward rapidity as a function of self-normalized 
charged-particle multiplicity in $p$$+$$p$ collisions at 
$\sqrts=200$~GeV by PHENIX compared with the (a) LHCb and (b) ALICE 
results at $\sqrts=13$~TeV~\cite{ALICE:2022gpu}.  The PHENIX data is 
compared when charged particles are measured at (a) forward rapidity 
and (b) backward rapidity. In addition, the data are compared to 
Detroit tuned~\cite{Aguilar:2021sfa} ([cyan] curve), including MPI 
effects. The error bars (boxes) denote the statistical (systematic) 
uncertainties.
}
\end{figure*}

The self-normalized $J/\psi$ yields 
($N_{J/\psi}/\langle{N_{J/\psi}}\rangle$) as a function of 
self-normalized charged-particle multiplicity (\nchach) are shown in 
Fig.~\ref{fig:jpsi_before_after} and compared to 
STAR~\cite{STAR:2018smh} and ALICE~\cite{ALICE:2021zkd,ALICE:2020msa} 
data.  The $J/\psi$ decay-daughter-muon contributions to the 
charged-particle multiplicity have been subtracted in the PHENIX 
results to remove the auto-correlation effect. Before subtracting the 
$J/\psi$ daughter-muon contributions, The PHENIX results show a 
multiplicity dependence that is similar to the STAR results at 
midrapidity. For ALICE at 13 TeV, the results at forward rapidity are 
systematically lower than at midrapidity, but it is difficult to reach 
conclusions for PHENIX and STAR at 200 GeV due to the larger systematic 
uncertainties.   After subtracting the \jpsi daughter-muon contributions, 
the \jpsi yields shift to lower \nchach, leading to a significant drop 
of $N_{J/\psi}/\langle N_{J/\psi}\rangle$, which is consistent with the 
results of opposite arms, where \nch is free from the \jpsi 
daughter-muon contributions. The PHENIX results with subtraction is 
significantly lower than the ALICE results at forward rapidity, but a 
comparable slope is seen in STAR and ALICE results at midrapidity. Note 
that the auto-correlation correction was not applied to the ALICE and 
STAR results.  The impact on the slope is expected to be more 
significant for the STAR than the ALICE results due to the smaller 
multiplicity.

The self-normalized $J/\psi$ yields are also compared with \pythia 
calculations. In Fig.~\ref{fig:jpsi_pythia} where the simulations are 
compared to $J/\psi$ normalized yield in data, the \pythia Detroit 
tuned with MPI calculations~\cite{Aguilar:2021sfa} agree with the data 
within $\approx{1}\sigma$ while failing to describe the data without 
MPI effects.  Therefore, MPI effects need to be included to accurately 
model \jpsi production in \pp collisions at RHIC. In addition, 
comparison between the PHENIX and ALICE results at forward rapidity 
indicates that the MPI effects on the multiplicity dependence would be 
even stronger at LHC energies.

The $(N_{\psi(2S)}/N_{J/\psi})/\langle N_{\psi(2S)}/N_{J/\psi}\rangle$ as a 
function of \nchach, shown in Fig.~\ref{fig:psip_alice_pythia}, was measured 
to investigate final-state interaction effects on charmonium production.  
The self-normalized yield ratios of \psipp to \jpsi at forward rapidity are 
shown as a function of normalized charged-particle multiplicity, where the 
charged-particle multiplicity has been measured using both the forward 
(same arms) and backward (opposite arms) FVTX detectors. The 
$({N_{\psi(2S)}/ N_{J/\psi})/\langle N_{\psi(2S)}/ N_{J/\psi}\rangle}$ 
measurements are consistent with unity within$\approx{1}\sigma$, and no 
significant rapidity dependence is observed. Therefore, final-state 
interactions on charmonium production appear negligible at the measured 
multiplicity range and collision energy.

Furthermore, in Fig.~\ref{fig:psip_alice_pythia}, the ALICE forward 
rapidity prompt \jpsi~\cite{ALICE:2022gpu} and LHCb prompt and 
nonprompt \jpsi measurements are included for comparison. The ALICE 
results also demonstrate similar consistency near unity up to 
$\NCH/\langle\NCH\rangle=\approx{6}$. Recently, LHCb has reported 
suppression of 
$(N_{\psi(1S)}/N_{J/\psi})/{\langle}N_{\psi(2S)}/N_{J/\psi}{\rangle}$ 
in high multiplicity \pp collisions~\cite{LHCb:2023xie}. The final 
state comover effect~\cite{Ferreiro:2014bia} becomes more pronounced at 
high multiplicity, and may result in the increased breakup of \psipp 
compared to \jpsi due to its lower binding energy.

Finally, Fig.~\ref{fig:psip_alice_pythia} also shows \pythia Detroit 
tuned with MPI effects for the 
($N_\psi(2S)/N_{J/\psi})/\langle{N_\psi(2S)}/N_{J/\psi}\rangle$ as a 
function of \nchach but without any final-state effects on quarkonia. 
The \pythia calculations yield similar results within uncertainties, 
and generally reproduce the data near unity. This indicates that the 
data favors no significant final-state interaction in \pp collisions at 
$\sqrts=200$~GeV.

%%%%%%%%%%%%%%%%%%%%%%%%%%%%%%%%%%%%%%%%%%%%%%%%%%%%%%%%%%%%%%%%%%
%%%%%%%%%%%%%%%%%%%%% Summary %%%%%%%%%%%%%%%%%%%%%%%%%%%%%%%%%
%%%%%%%%%%%%%%%%%%%%%%%%%%%%%%%%%%%%%%%%%%%%%%%%%%%%%%%%%%%%%%%%%%
% Summary

In summary, we have reported multiplicity-dependent studies of \jpsi 
and \psipp production at forward rapidity in \pp collisions at 
$\sqrts=200$~GeV. The self-normalized \jpsi yields are measured up to 
$\approx$6 units of \nchach and reasonable consistency with the STAR 
and ALICE data has been achieved when the \jpsi and charged particles 
are measured in the same pseudorapidity region. Results of 
self-normalized \jpsi yields before and after \jpsi decay-daughter-muon 
subtraction for \nch counting can be well described by the \pythia 
Detroit tuned with MPI. Also, forward self-normalized \psipp to \jpsi 
yields as a function of forward and backward track multiplicity have 
been studied. Agreements with unity have been observed within 
uncertainties and the results are consistent with each other at 
different rapidity gaps, suggesting no significant final-state 
interactions on charmonium production. The \pythia model without 
final-state interactions generally reproduces the self-normalized 
\psipp to \jpsi yields indicating that no such effect occurs in \pp 
collisions at RHIC energies. 

Consistency with unity is also reported by ALICE. However, LHCb 
observes a suppression of prompt $\psi(2S)$ production at high 
multiplicity, aligning with the prediction of final-state comover 
effects~\cite{FERREIRO201457}. The LHCb results also demonstrate 
consistency with the PHENIX data points within uncertainties. 
Investigations of charmonium production in $p$$+$Al and $p$$+$Au 
collisions with PHENIX data, particularly in the backward-rapidity 
region, would provide an excellent opportunity to further study 
final-state effects in small collision systems at RHIC energies.

%%%%%%%%%%%%%%%%%%%%%%%%%%%%%%%%%%%%%%%%%%%%%%%%%%%%%%%%%%%%%%%%%%
%%%%%%%%%%%%%%%%%%%%% Acknowledgements %%%%%%%%%%%%%%%%%%%%%%%%%%%
%%%%%%%%%%%%%%%%%%%%%%%%%%%%%%%%%%%%%%%%%%%%%%%%%%%%%%%%%%%%%%%%%%

%%%%%%%%%%%%%%%%%%%%%%  ACKNOWLEDGMENTS}  %%%%% MGS22a version
%% 2018 change in Korea
%%% 2021 change in dropping Brazil, Germany, and Pakistan, because
%%%      they no longer have active MGS and left PHENIX before 2015
%% 2024 add HUN-REN ATOMKI [and remove some] (Hungary)

%\begin{acknowledgments}

We thank the staff of the Collider-Accelerator and Physics
Departments at Brookhaven National Laboratory and the staff of
the other PHENIX participating institutions for their vital
contributions.  
We acknowledge support from the Office of Nuclear Physics in the
Office of Science of the Department of Energy,
the National Science Foundation,
Abilene Christian University Research Council,
Research Foundation of SUNY, and
Dean of the College of Arts and Sciences, Vanderbilt University
(U.S.A),
Ministry of Education, Culture, Sports, Science, and Technology
and the Japan Society for the Promotion of Science (Japan),
Conselho Nacional de Desenvolvimento Cient\'{\i}fico e
Tecnol{\'o}gico and Funda\c c{\~a}o de Amparo {\`a} Pesquisa do
Estado de S{\~a}o Paulo (Brazil),
Natural Science Foundation of China (People's Republic of China),
Croatian Science Foundation and
Ministry of Science and Education (Croatia),
Ministry of Education, Youth and Sports (Czech Republic),
Centre National de la Recherche Scientifique, Commissariat
{\`a} l'{\'E}nergie Atomique, and Institut National de Physique
Nucl{\'e}aire et de Physique des Particules (France),
J. Bolyai Research Scholarship, EFOP, HUN-REN ATOMKI, NKFIH,
MATE KKF, and OTKA (Hungary), 
Department of Atomic Energy and Department of Science and Technology (India),
Israel Science Foundation (Israel),
Basic Science Research and SRC(CENuM) Programs through NRF
funded by the Ministry of Education and the Ministry of
Science and ICT (Korea).
Ministry of Education and Science, Russian Academy of Sciences,
Federal Agency of Atomic Energy (Russia),
VR and Wallenberg Foundation (Sweden),
University of Zambia, the Government of the Republic of Zambia (Zambia),
the U.S. Civilian Research and Development Foundation for the
Independent States of the Former Soviet Union,
the Hungarian American Enterprise Scholarship Fund,
the US-Hungarian Fulbright Foundation,
and the US-Israel Binational Science Foundation.

%\end{acknowledgments}

%\textbf{*** page break for PRD Letters word count $<$4.5 pages $<$9 columns} 
%\clearpage

%\section{APPENDIX:DATA TABLES}

%%%%%%%%%%%%%%%%%%%%%%%%%%%  References

%\bibliography{ppg256x0}

 
\end{document}